\documentclass[a4paper,12pt, epsfig]{article}
\def\letter{0}\def\pr{0}
\pdfoutput=1 %per jhep
\usepackage{epsfig}
\usepackage{epstopdf}
\usepackage{graphicx}
\usepackage{ifthen}
\usepackage{amsmath}
\usepackage[psamsfonts]{amssymb}
\usepackage{euscript}

\usepackage{latexsym}
\usepackage[arrow,matrix,curve]{xy}

\usepackage[hypertexnames=false]{hyperref}
\hypersetup{
    colorlinks=true,
    linkcolor=blue,
    filecolor=magenta,   
    citecolor=black,   
    urlcolor=cyan,
    pdftitle={Overleaf Example},
    pdfpagemode=FullScreen,
    }

\newskip\humongous \humongous=0pt plus 1000pt minus 1000pt

\newif\ifdtup

\def\,{\hspace{-.1cm}}
\def\hsp{,\hspace{.7cm}}

\def\tf {\tilde{f}}
\def\ff{{\mathcal{F}}}

\def\fc#1#2 {\frac{n}{q}#1\frac{n}{q}#2}

\def\hf{H\p_{f_0}}
\def\hpt{H_{\rm{free}}}

\newcommand{\vac}{\ensuremath{|0\rangle}}
\newcommand{\stt}{\ensuremath{|\psi\rangle}}

\renewcommand{\cos}{\textrm{cos}}
\renewcommand{\sin}{\textrm{sin}}

\renewcommand{\sinh}{\textrm{sinh}}
\renewcommand{\cosh}{\textrm{cosh}}
\renewcommand{\tanh}{\textrm{tanh}}
\newcommand{\sech}{\textrm{sech}}
\newcommand{\csch}{\textrm{csch}}

\def\exp#1{\hbox{\rm exp}\left(#1\right)}

\renewcommand{\theequation}{\arabic{section}.\arabic{equation}}
\renewcommand{\(}{\begin{equation}}
\renewcommand{\)}{end{equation} \vspace{-.05in}\linebreak}

\newcounter{saveeqn}
\newcounter{savealpheqn}

\newcommand{\alpheqn}{\setcounter{saveeqn}{\value{equation}}%
  \stepcounter{saveeqn}\setcounter{equation}{0}%
  \renewcommand{\theequation}{\mbox{\arabic{section}.\arabic{saveeqn}
\alph{equation}}}
  \renewcommand{\)}{\end{equation}}}
\def\part#1{\frac{\partial}{\partial{#1}}}%
\def\group#1{\refstepcounter{equation}\setcounter{saveeqn}
 {\value{equation}}%
  \label{#1}\setcounter{equation}{0}%
\renewcommand{\theequation}{\mbox{\arabic{section}.\arabic{saveeqn}
\alph{equation}}}
  \renewcommand{\)}{\end{equation}}}
\newcommand{\reseteqn}{\setcounter{equation}{\value{saveeqn}}%
  \renewcommand{\theequation}{\arabic{section}.\arabic{equation}}%
  \renewcommand{\)}{\end{equation}}}

\newcommand{\aalpheqn}{\setcounter{saveeqn}{\value{equation}}%
  \stepcounter{saveeqn}\setcounter{equation}{0}%
  \renewcommand{\theequation}{\mbox{
        \Alph{subsection}.\arabic{saveeqn}\alph{equation}}}
   \renewcommand{\)}{\end{equation}}}
\newcommand{\areseteqn}{\setcounter{equation}{\value{saveeqn}}%
  \renewcommand{\theequation}{\Alph{subsection}.\arabic{equation}}%
  \renewcommand{\)}{\end{equation}}}

\renewcommand{\thefootnote}{\alph{footnote}}
\renewcommand{\(}{\begin{equation}}
\renewcommand{\)}{\end{equation}}
\newcommand{\ba}{\begin{eqnarray}}
\newcommand{\ea}{\end{eqnarray}}

\renewcommand{\b}{\beta}

\newcommand{\vt}{V^{(3)}(\sqrt{\lambda}f(x))}
\newcommand{\cbp}{\mathop{\vtop{\ialign{##\crcr
   $\hfil\displaystyle{}\hfil$\crcr\noalign{\kern-13pt\nointerlineskip}
   \BIG{)}\hskip0pt\crcr\noalign{\kern3pt}}}}}
\newcommand{\pa}{\mathop{\vtop{\ialign{##\crcr

$\hfil\displaystyle{\oplus}\hfil$\crcr\noalign{\kern+1pt\nointerlineskip
}
   \hspace{.08in}$^{\alpha=0}$\hskip6pt\crcr\noalign{\kern3pt}}}}}
\renewcommand{\hsp}{,\hspace{.3in}}
\newcommand{\p}{^\prime}
\newcommand{\pp}{^{\prime\prime}}

\def\D{\ensuremath{{\cal D}}}
\catcode`\@=11
\def\vereq#1#2{\lower3pt\vbox{\baselineskip1.5pt \lineskip1.5pt
\ialign{$\m@th#1\hfill##\hfil$\crcr#2\crcr\sim\crcr}}}
\catcode`\@=12

\renewcommand{\(}{\begin{equation}}
\renewcommand{\)}{\end{equation}}

\def\pin#1{\int \frac{d#1}{2\pi}}
\def\ppin#1{\int\hspace{-17pt}\sum \frac{d#1}{2\pi}}
\def\ppink#1{\int\hspace{-17pt}\sum\frac{d^{#1}k}{(2\pi)^{#1}}}
\def\dint{\int\hspace{-12pt}\sum }
\def\pink#1{\int \frac{d^{#1}k}{(2\pi)^{#1}}}

\def\Bd#1{B^\dag_{k_{#1}}}

\def\tp#1#2#3{\hbox{\rm tan}^#1(\thet#2#3)}

\def\cc{\mathcal{C}}
\def\df{\mathcal{D}_{f}}

\def\dF{\mathcal{D}_F}

\def\I{\mathcal{I}}

\def\os{\omega_S}

\def\as{|\alpha;\sigma\rangle}
\def\asb{\langle\alpha;\sigma|}

\newcommand{\beas}{\begin{eqnarray*}}
\newcommand{\eeas}{\end{eqnarray*}}

\newcommand{\bquo}{\begin{quote}}
\newcommand{\enqu}{\end{quote}}

\newcommand{\g}{{\mathfrak g}}

\def\ch{{\mathcal{H}}}

\def\tp{{\tilde{\phi}}}
\def\ok#1{\omega_{k_{#1}}}

\def\V#1{V^{(#1)}(gf(x))}

\def\ck{\csch\left(\frac{\pi k}{2\b}\right)}

\newcommand{\beq}{\begin{equation}}
\newcommand{\eeq}{\end{equation}}
\newcommand{\bea}{\begin{eqnarray}}
\newcommand{\eea}{\end{eqnarray}}

\newskip\humongous \humongous=0pt plus 1000pt minus 1000pt

\newif\ifdtup

\catcode`\@=11

\ifthenelse{\equal{\letter}{0}}{ %se lettere=0, commenta questo

\@addtoreset{equation}{section}
\def\theequation{\arabic{section}.\arabic{equation}}

\def\@normalsize{\@setsize\normalsize{15pt}\xiipt\@xiipt
\abovedisplayskip 14pt plus3pt minus3pt%
\belowdisplayskip \abovedisplayskip
\abovedisplayshortskip \z@ plus3pt%
\belowdisplayshortskip 7pt plus3.5pt minus0pt}

\def\small{\@setsize\small{13.6pt}\xipt\@xipt
\abovedisplayskip 13pt plus3pt minus3pt%
\belowdisplayskip \abovedisplayskip
\abovedisplayshortskip \z@ plus3pt%
\belowdisplayshortskip 7pt plus3.5pt minus0pt
\def\@listi{\parsep 4.5pt plus 2pt minus 1pt
      \itemsep \parsep
      \topsep 9pt plus 3pt minus 3pt}}

\relax

\def\section{\@startsection{section}{1}{\z@}{3.5ex plus 1ex minus  .2ex}{2.3ex plus .2ex}{\large\bf}}

\def\thesection{\arabic{section}}
\def\thesubsection{\arabic{section}.\arabic{subsection}}

\def\appendix{\setcounter{section}{0}
 \def\thesection{Appendix \Alph{section}}
 \def\thesubsection{\Alph{section}.\arabic{subsection}}
 \def\theequation{\Alph{section}.\arabic{equation}}}
\renewcommand{\theequation}{\arabic{section}.\arabic{equation}}

}{
\renewcommand{\theequation}{\arabic{equation}}

} %se letter=0, commenta questo

\begin{document}
% ========================================================================
\def\thefootnote{\fnsymbol{footnote}}
\def\thetitle{Quantum Reflective Kinks}
\def\autone{Jarah Evslin}
\def\auttwo{Hui Liu}
\def\affa{Institute of Modern Physics, NanChangLu 509, Lanzhou 730000, China}
\def\affb{University of the Chinese Academy of Sciences, YuQuanLu 19A, Beijing 100049, China}
\def\affc{School of Fundamental Physics and Mathematical Sciences, Hangzhou Institute for Advanced Study,
University of Chinese Academy of Sciences, Hangzhou 310024, China}
\def\affd{Arnold Sommerfeld Center, Ludwig-Maximilians-Universität, Theresienstraße 37, 80333 München, Germany}

\title{Titolo}

\ifthenelse{\equal{\pr}{1}}{
\title{\thetitle}
\author{\autone}
\author{\auttwo}
\affiliation {\affa}
\affiliation {\affb}
%\affiliation {\affc}
%\affiliation {\affd}
%pr e uno

}{

\begin{center}
{\large {\bf \thetitle}}

\bigskip

\bigskip

%\catcode`@=11

{\large \noindent  \autone{${}^{1,2}$} \footnote{jarah@impcas.ac.cn} %}
and \auttwo{${}^{3,2,4}$} \footnote{hui.liu@campus.lmu.de}}

%{\large \noindent  \autone{${}^{1,2}$} \footnote{jarah@impcas.ac.cn} and \auttwo{${}^{1,2}$} \footnote{guohengyuan@impcas.ac.cn}}

\vskip.7cm

1) \affa\\
2) \affb\\
3) \affc\\
4) \affd\\

\end{center}

}

\begin{abstract}
\noindent
We scatter a meson off of a scalar kink in quantum field theory, at leading order in perturbation theory.  We calculate the full quantum state, at leading order, at all times and also check that the reflection and transmission coefficients agree with those which would be obtained in relativistic quantum mechanics.

\end{abstract}

% \vfill
%
% \end{titlepage}
\setcounter{footnote}{0}
\renewcommand{\thefootnote}{\arabic{footnote}}

\ifthenelse{\equal{\pr}{1}}
{
\maketitle
}{}

\section{Introduction}
Classical kink-(anti)kink scattering has been a major industry since the discovery of a fractal pattern of resonance windows in the $\phi^4$ double well model in Ref.~\cite{csw}.  For example, there has recently been a flurry of activity trying to understand whether the resonances are caused by the shape modes of individual kinks or collective bound modes of the combined kink-antikink system \cite{doreynoshape,doreyinterp,tomcc}.

In contrast, the much simpler process of kink-meson scattering has received relatively little attention \cite{faddeev77,lowe79,parm87,swanson88,uehara91,abdel11}.  In classical $\phi^4$ field theory, Refs.~\cite{tomrad1,tomrad2} discovered that the mesons apply a negative pressure to the kinks.  The reason for this negative pressure is quite simple.  Inside of the kink, $n$ mesons of frequency $\omega$ can fuse into one meson of frequency $n\omega$, which as a result of the mass shell condition has more momentum than the sum of the initial $n$ mesons.  Conservation of momentum demands that this excess momentum comes from the kink, forcing the kink to move in the opposite direction.   

This meson fusion appears to be the only phenomenon which occurs in meson-kink scattering in the classical $\phi^4$ model.  However in the case of reflective kinks, the mesons may also reflect, leading to a positive pressure \cite{tomrad3}.  Thus, in general, two processes are allowed in classical kink-meson scattering, fusion and reflection.

In the quantum theory, we expect a much richer phenomenology.  For one, an infinite tower of unstable excited states corresponding to multiple shape mode excitations is expected.  These should appear as narrow resonances in elastic kink-meson scattering.   Inelastic kink-meson scattering is also expected to be rich.   For example, Raman spectroscopy will be possible, in which a monochromatic meson scatters off of the kink, exciting an internal shape mode.  The kink's excitation spectrum can be read off of the observed energy decreases of the scattered meson.  Deexcitation of a shape mode is also possible, in the inverse process.  Finally, in the quantum theory, one expects not only meson fusion, but also meson fission.

All of these processes are expected to occur at tree level in the kink Hamiltonian, introduced in Ref.~\cite{dhn2}.  However, in traditional approaches which allow multiloop calculations, such as the collective coordinate method of Refs.~\cite{gjscc,gj76}, they have been prohibitively difficult to calculate.  Recently a new approach, linearized kink perturbation theory, has been introduced at one-loop in Ref.~\cite{mekink} and beyond in Ref.~\cite{me2loop}.  This new formalism is particularly simple\footnote{It requires a choice of base point in moduli space.  If the kink moves too far from the base point, as will happen in kink-(anti)kink scattering, then one must compose the evolutions calculated at distinct base points.} in the one-kink sector, which contains a single kink and any number of mesons and impurities.  It is thus well suited to the scattering of a meson with a nonrelativistic kink.

We plan to investigate all of these new tree-level phenomena in the near future.  However, so far applications of the formalism have suffered from two limitations.  First, all explicit calculations have considered only kinks described by reflectionless potentials\footnote{An exception to this is Ref.~\cite{chris}, however the treatment of the particular features of reflective normal modes was handled inside of the code that performed the numerical integrals, and only described roughly in an Appendix.  The treatment here is related to that which was used inside the numerical code.} and second, all calculations so far have been time-independent.  

In the present note, we treat kink-meson scattering using the free kink Hamiltonian.   As will be reviewed below, the free kink Hamiltonian describes a free theory in the sense that it is a sum of quantum harmonic oscillators, but it has a position-dependent mass term which leads to some dynamics.  Clearly, this is a calculation which needs to be performed before our general study of tree level interactions.  In doing so, we treat the two shortcomings noted above.  In particular, we treat not only reflectionless but also reflective kinks.  We evolve the system in time, and so we observe that our treatment of reflective kinks, in the free theory, reproduces the same reflection and transmission coefficients that would be calculated in relativistic quantum mechanics.

%If the potential is not reflectionless, then a meson can elastically scatter off of a kink even at $O(\lambda^0)$.  This process is described by evolution under the free kink Hamiltonian $H_2$.

\section{Review}

Consider a (1+1)-dimensional Schrodinger picture quantum field theory of a scalar meson field $\phi(x)$ with conjugate $\pi(x)$ defined by the Hamiltonian $H$
\bea
\ch(x)&=&\frac{1}{2}:\pi(x)\pi(x):_a+\frac{1}{2}:\partial_x\phi(x)\partial_x\phi(x):_a+\frac{1}{g^2}:V(g\phi(x)):_a\nonumber\\
H&=&\int dx \ch(x)\hsp 
m^2=V^{(2)}(gf(\pm\infty)). \label{hd}
\eea
Here $::_a$ is the usual normal-ordering with mass $m$, $\V{n}$ is the $n$th derivative of $V$ with respect to its argument and $f(x)$ is a kink solution
\beq
\phi(x,t)=f(x)\hsp
-gf^{''}(x)+\V{1}=0.
\label{fd}
\eeq
If the two sign choices in the definition of $m^2$ lead to different values, then the quantum-corrected vacuum energies will differ on the two sides of the kink and so the kink will accelerate \cite{weigelstab}.  In this case the kink states are never Hamiltonian eigenstates and so we will not consider this case further.

In terms of the defining Hamiltonian, the kink states are nonperturbative.  In classical field theory this corresponds to the fact that the classical field $\phi(x)$ is far from a minimum of the potential, and it would be fixed by writing the Hamiltonian in terms of $\phi(x)-f(x)$.  In the quantum theory such a replacement is ill-defined as a result of the regularization, which in this case is achieved via normal ordering.  Indeed a naive application of the replacement can change the spectrum of the Hamiltonian \cite{rebhan} and so lead to wrong answers.  To ensure that the spectrum of the Hamiltonian is not changed, we transform it to the kink Hamiltonian $H\p$ by conjugating with a unitary displacement operator
\beq
\df={{\rm Exp}}\left[-i\int dx f(x)\pi(x)\right]\hsp
H\p=\df^\dag H\df. \label{df}
\eeq
This displacement operator plays the same role in the quantum theory as the shift in the 
classical theory
\beq
:F[\phi(x),\pi(x)]:_a\df=\df:F[\phi(x)+f(x),\pi(x)]:_a.
\eeq

In summary, we have constructed a kink Hamiltonian $H\p$ which has the same spectrum as the defining Hamiltonian $H$.  While the defining Hamiltonian generates time evolution in the defining frame of the Hilbert space, the unitary transformation $\df^\dag$ takes this defining frame to the kink frame.  Masses may be measured and time may be evolved in the kink frame using $H\p$.  The isospectral property means that these masses will be the correct ones
\beq
H|K\rangle=E|K\rangle \Rightarrow H\p \df^\dag |K\rangle = E \df^\dag|K\rangle
\eeq
and an application of $\df$ after time evolution yields the time evolution that would be calculated using the defining Hamiltonian \cite{rob}
\beq
\df e^{-iH\p t}\df^\dag=e^{-iHt}.
\eeq
Thus nonperturbative problems using $H$ can be transformed into problems which, in the one-kink sector, are perturbative using $H\p$.

One can use Eq.~(\ref{df}) to calculate the kink Hamiltonian $H\p$.  Let us decompose it into terms $H\p_j$ which, when normal ordered, have $j$ factors of $\phi$ and its conjugate $\pi$.  These terms will be of order $g^{j-2}$.  Then $H\p_0$ is the classical kink mass $Q_0$, while the tadpole $H\p_1$ vanishes as a result of the classical equations of motion (\ref{fd}).  The first nontrivial term is the free kink Hamiltonian
\beq
H\p_2=\frac{1}{2}\int dx\left[:\pi^2(x):_a+:\left(\partial_x\phi(x)\right)^2:_a+\V{2}:\phi^2(x):_a\right]. \label{h2}
\eeq
It is free, in the sense that it is quadratic in the fields, but notice that the mass term depends on $x$.  This means that the constant frequency solutions of its classical equations of motion
\beq
\V{2}{\g}(x)=\omega^2{\g}(x)+{\g}^{\prime\prime}(x)\hsp \phi(x,t)=e^{-i\omega t}\g(x) \label{sl}
\eeq
are not plane waves, but rather are normal modes.

In general there are three kinds of normal mode, classified by their frequencies $\omega$.  There is always a zero mode $\g_B(x)=f\p(x)/\sqrt{Q_0}$ with zero frequency $\omega_B=0$.  Sometimes there will be real shape modes $\g_{S}(x)$ with frequencies $0<\omega_{S}<m$.  For each real $k$ there will be a continuum mode with frequency $\omega_k$
\beq
\ok{}=\sqrt{m^2+k^2}.
\eeq
As $\omega_k=\omega_{-k}$ one needs to define the decomposition into these two modes.  We will require
\beq
{\g}_{-k}(x)={\g}_k^*(x)
\eeq
and we will fix the normalization and decomposition of all modes using the completeness relations\footnote{Completeness follows from the fact that (\ref{sl}) is a Sturm-Liouville equation.} in $k$ space
\beq
\int dx |{\g}_{B}(x)|^2=1,\
\int dx {\g}_{k_1} (x) {\g}^*_{k_2}(x)=2\pi \delta(k_1-k_2),\ 
\int dx {\g}_{S_1}(x){\g}^*_{S_2}(x)=\delta_{S_1S_2} \label{comp}
\eeq
and in $x$ space
\beq
{\g}_B(x){\g}_B(y)+\ppin{k}{\g}_k(x){\g}^*_{k}(y)=\delta(x-y)\hsp
\ppin{k}=\pin{k}+\sum_S \label{compx}
\eeq
where we have defined $\dint$ to be an integral over $k$ together with a sum over shape modes $S$.

As the normal modes are a complete basis of bounded functions of $x$, and we are working in the Schrodinger picture where the fields only depend on $x$, we may follow Ref.~\cite{cahill76} and decompose the fields in terms of normal modes
\bea
\phi(x)&=&\phi_0 {\g}_B(x) +\ppin{k}\left(B_k^\ddag+\frac{B_{-k}}{2\omega_k}\right) {\g}_k(x)\label{bdec}\\
\pi(x)&=&\pi_0 {\g}_B(x)+i\ppin{k}\left(\omega_kB_k^\ddag - \frac{B_{-k}}{2}\right) {\g}_k(x)\nonumber
\eea
where $B^\ddag_k=B^\dag_k/(2\ok{})$ and $B_{-S}=B_{S}$.  Thus any operator in the theory may be expanded in terms of the operators $\phi_0,\ \pi_0,\ B_S,\ B^\ddag_S,\ B_k$ and $B^\ddag_k$ which, as a result of the canonical commutation relations satisfied by $\phi(x)$ and $\pi(x)$, satisfy the algebra
\beq
{[\phi_0,\pi_0]}=i\hsp
[B_{S_1},B^\ddag_{S_2}]=\delta_{S_1S_2}\hsp
[B_{k_1},B^\ddag_{k_2}]=2\pi\delta(k_1-k_2).\nonumber
\eeq

Inserting the decomposition (\ref{bdec}) into the formula (\ref{h2}) for the free kink Hamiltonian, we find an enormous simplification
\beq
H\p_2=Q_1+\frac{\pi_0^2}{2}+\ppin{k}\omega_k B^\ddag_k B_k \label{h2a}
\eeq
where
\beq
Q_1=-\frac{1}{4}\ppin{k}\pin{p}\frac{(\omega_p-\omega_k)^2}{\omega_p}\tilde{{\g}}^2_{k}(p)-\frac{1}{4}\pin{p}\omega_p\tilde{{\g}}_{B}(p)\tilde{{\g}}_{B}(p)\hsp
\tilde{{\g}}(p)=\int dx {\g}(x) e^{ipx}.
\eeq
We have found that the free kink Hamiltonian $H\p_2$ is the sum of three terms.  The first, $Q_1$, is the one-loop kink mass in the form found in Ref.~\cite{cahill76}.  The second describes the quantum mechanics of a free particle, in this case the center of mass of the kink.  The third is an infinite sum of quantum harmonic oscillators, one for each continuum mode $\g_k(x)$ and also one for each shape mode $\g_S(x)$ if there are any.  The ground state $\vac_0$ of the free kink Hamiltonian therefore corresponds to the ground state of each of these quantum mechanical systems
\beq
\pi_0\vac_0=B_k\vac_0=B_S\vac_0=0. \label{v0}
\eeq
This state is the first approximation in a semiclassical expansion of the kink ground state, in the kink frame.   At this order, continuum normal modes and shape modes can be excited in the kink frame by acting with $B^\ddag_k$ and $B^\ddag_S$ respectively.  We remind the reader that the corresponding states in the defining frame of the Hilbert space are then obtained by acting with $\df$, so that $\df\vac_0$ is the leading term in the kink ground state.

\section{Normal Modes}

Consider normal modes with asymptotic behavior
\bea
\g_k(x)&=&\left\{\begin{tabular}{lll}
$B_ke^{ikx}+C_k e^{-ikx}$&\rm{if} & $x\ll  -1/m$\\
$D_ke^{ikx}+E_k e^{-ikx}$&\rm{if} & $x\gg 1/m$\\
\end{tabular}
\right. \label{gk}\\
B^*_k&=&B_{-k}\hsp
C^*_k=C_{-k}\hsp
D^*_k=D_{-k}\hsp
E^*_k=E_{-k}.\nonumber
\eea
The coefficients $B$, $C$, $D$ and $E$ are constrained by the completeness relations (\ref{comp}) and (\ref{compx}).  Normal modes at distinct values of $|k|$ will automatically be orthogonal as they satisfy the same Sturm-Liouville equation with a distinct eigenvalue.  However some care needs to be taken to ensure that $\g_k(x)$ and $\g_{-k}(x)$ are orthogonal and correctly normalized.   As the nonvanishing integrals are infinite, we only consider the large $|x|$ region.  %There are two completeness relations to impose, one involving $x$ integration and the other $k$ integration.

%The first completeness relation is
%\beq
%\int_{-\infty}^\infty dx g_{k_1}(x) g_{k_2}(x)=2\pi\delta(k_1+k_2).
%\eeq
Let us impose the $k$-space completeness relation (\ref{comp}) for the modes $\g_k$.  Since we are only interested in large $|x|$, it is sufficient to impose
\beq
\stackrel{\rm{lim}}{{}_{L\rightarrow\infty}}\left[\int_{x=-2L}^{x=-L}+\int_{x=L}^{x=2L}\right]dx \g_{k_1}(x) \g_{k_2}(x)=
\left\{\begin{tabular}{lll}
$2L$&\rm{if} & $k_1=-k_2$\\
$0$&\rm{if} & $k_1\neq -k_2$.\\
\end{tabular}\right.
\eeq
This is easily evaluated
\beq
\stackrel{\rm{lim}}{{}_{L\rightarrow\infty}}\left[\int_{x=-2L}^{x=-L}+\int_{x=L}^{x=2L}\right]\frac{dx}{L} \g_{k_1}(x) \g_{k_2}(x)=
\left\{\begin{tabular}{lll}
$|B_k|^2+|C_k|^2+|D_k|^2+|E_k|^2$&\rm{if} & $k=k_1=-k_2$\\
$2B_{k}C_{k}+2D_{k}E_{k}$&\rm{if} & $k=k_1=k_2$\\
$0$&\rm{if} & $k_1\neq \pm k_2$.\\
\end{tabular}\right.
\eeq
Then we learn that
\beq
|B_k|^2+|C_k|^2+|D_k|^2+|E_k|^2=2\hsp
B_{k}C_{k}+D_{k}E_{k}=0. \label{ortog}
\eeq

%The other completeness relation is
%\beq
%\g_B(x)\g_B(y)+\ppin{k} \g_k(x) \g_{-k}(y)=\delta(x-y).
%\eeq
Next we impose the completeness relation in position space (\ref{compx}).  When $|x|,\ |y|\gg 1/m$ only the continuum modes contribute, leaving
\beq
\pin{k} \g_k(x) \g_{-k}(y)=\delta(x-y).
\eeq
Evaluating the left hand side we find
\beq
\pin{k} \g_k(x) \g_{-k}(y)=\pin{k}\left\{\begin{tabular}{lll}
$\left(|B_k|^2+|C_k|^2\right)e^{ik(x-y)}$&\rm{if} & $x,\ y\ll -1/m$\\
$\left(|D_k|^2+|E_k|^2\right)e^{ik(x-y)}$&\rm{if} & $x,\ y\gg 1/m$\\
$\left(B^*_kE_k+C_kD^*_k\right)e^{-ik(x+y)}$&\rm{if} & $x\ll -1/m,\ 1/m\ll  y$.\\
\end{tabular}\right.
\eeq
We are interested in the case in which $|x|$ and $|y|$ are very large and so $B_k$, $C_k$ and $D_k$ and $E_k$ vary much more slowly than the plane wave factors with respect to $k$.  To avoid an unwanted contribution at $x=-y$ we impose the constraint
\beq
B^*_kE_k+C_kD^*_k=0 \label{ort}.
\eeq
In addition, to arrive at the correct normalization of the delta function in the limit of large $|x|$ and $|y|$, one finds the constraints
\beq
|B_k|^2+|C_k|^2=|D_k|^2+|E_k|^2=1
. \label{unid}
\eeq
We can summarize all of these constraints with the condition that the matrix
\beq
U=\left(
\begin{tabular}{cc}
$B_k$&$C^*_k$\\
$E_k$&$D^*_k$\
\end{tabular}
\right)\hsp
U^\dag U=1 \label{un}
\eeq
is unitary.

\section{Propagating Wave Packets}

Consider a moving wave packet
\beq
\Phi(x)={\rm Exp}\left[-\frac{(x-x_0)^2}{4\sigma^2}+i x k_0\right] \hsp 
x_0\ll -1/m\hsp
k_0\gg \frac{1}{\sigma}\hsp
\sigma\ll |x_0|. \label{wp}
\eeq
This corresponds to a meson beginning far to the left of a kink, at $x=x_0$, and moving to the right with momentum $k_0$.  We will assume that it moves fast enough that smearing can be neglected.

The completeness of the normal modes implies that, at $|x|\gg 1/m$, any wave packet may be decomposed
\beq
\Phi(x)=\pin{k} \alpha_k \g_k(x)\hsp \alpha_k=\int dx \Phi(x) \g^*_k(x).
\eeq
The wave packet (\ref{wp}) is supported at $x\ll -1/m$ and so one may insert the asymptotic formula for $\g^*_k(x)$ valid at $x\ll -1/m$.  Therefore
\beq
\alpha_k=2\sigma\sqrt{\pi}\left(
B^*_k e^{-i(k-k_0)x_0}e^{-(k-k_0)^2\sigma^2}+C^*_k e^{i(k+k_0)x_0}e^{-(k+k_0)^2\sigma^2} 
\right). \label{alf}
\eeq
In Dirac notation, in the kink frame, the wave packet corresponds to the one-meson state 
\beq
|\Phi\rangle=\int dx \Phi(x)|x\rangle=\pin{k}  \alpha_k|k\rangle\hsp |k\rangle=B^\ddag_k\vac_0\hsp |x\rangle=\pin{k}\g^*_k(x)|k\rangle \label{fock}
\eeq
which is part of the meson Fock space decomposition of the one-kink sector.   Thus we begin with a meson at $x=x_0$ moving towards a kink at $x\sim 0$.  We remind the reader that $\vac_0$, defined in Eq.~(\ref{v0}), is the lowest order approximation to the ground state of a single kink, as expressed in the kink frame of the Hilbert space.

At time $t$ the wave packet evolves to
\beq
|\Phi(t)\rangle=\int dx \Phi(x,t)|x\rangle\hsp
\Phi(x,t)=\pin{k} e^{-i\omega_k t}\alpha_k \g_k(x). \label{ft}
\eeq
With $\sigma$ large enough, $k$ is always close to $k_0$ or $-k_0$.  Therefore we will expand in powers of $(k\pm k_0)$, out to first order.  Wave packet smearing appears only at second order and so will be missed by this approximation.  At first order
\beq
\omega_k=\omega_{k_0}+(\pm k - k_0)\frac{k_0}{\omega_{k_0}}. \label{om}
\eeq
Inserting (\ref{alf}) and (\ref{om}) into Eq.~(\ref{ft}) we find the wave packet evolution. 

At $x\ll 0$ we obtain the wave packet
\bea
\Phi(x,t)&=&2\sigma\sqrt{\pi}e^{-i\omega_{k_0}t}\pin{k} \left(
B^*_k  e^{-i(k-k_0)\left(x_0+\frac{k_0}{\omega_{k_0}}t\right)}e^{-(k-k_0)^2\sigma^2}\right.\nonumber\\
&&\left.+C^*_k e^{i(k+k_0)\left(x_0+\frac{k_0}{\omega_{k_0}}t\right)}e^{-(k+k_0)^2\sigma^2} 
\right) \left(B_ke^{ikx}+C_k e^{-ikx}\right).
\eea
For $\sigma$ large enough, the first Gaussian is supported at $k=k_0$  and the second at $k=-k_0$ and so we approximate the coefficients $B_k$ and $C_k$ by their values at $k=\pm k_0$, yielding
\bea
\Phi(x,t)&=&e^{-i\omega_{k_0}t}\left(
\left(\left|B_{k_0}\right|^2+\left|C_{k_0}\right|^2  \right){\rm Exp}\left[{-\frac{\left(-x+x_0+\frac{k_0}{\omega_{k_0}}t\right)^2}{4\sigma^2}}+ik_0 x\right]\right.\nonumber\\
&&
\left.+2B^*_{k_0}C_{k_0}{\rm Exp}\left[{-\frac{\left(x+x_0+\frac{k_0}{\omega_{k_0}}t\right)^2}{4\sigma^2}}-ik_0 x\right]
\right).
\eea

The two Gaussian factors are supported, respectively, when the position $x$ is equal to plus or minus
\beq
x_t=x_0+\frac{k_0}{\omega_{k_0}}t.
\eeq
Here we recognize $k_0/\omega_{k_0}$ as the group velocity of the wave packet and $t$ as the propagation time. We are now considering the case $x\ll 0$, and so the first Gaussian is only supported when $x_t\ll 0$ and the second when $x_t\gg 0$.  

Thus at early times $t\ll \omega_{k_0} |x_0|/k_0$, only the first Gaussian factor may contribute and we find
\beq
\Phi(x,t)=e^{-i\omega_{k_0}t+ik_0 x}{\rm Exp}\left[{-\frac{\left(-x+x_0+\frac{k_0}{\omega_{k_0}}t\right)^2}{4\sigma^2}}\right]=e^{-i\left(m^2/\omega_{k_0}\right)t}\Phi\left(x-\frac{k_0}{\omega_{k_0}}t\right)
\eeq
where we have used (\ref{unid}).  The interpretation is that, to this order in the $(k-k_0)$ expansion, the wave packet simply moves rigidly to the right before arriving at the kink.  At times $t\gg \omega_{k_0} |x_0|/k_0$, after the meson and kink have interacted, only the second Gaussian may contribute.  Thus, at $x\ll -1/m$, the wave function becomes
\beq
\Phi(x,t)=2e^{-i\omega_{k_0}t-ik_0 x}B^*_{k_0}C_{k_0}{\rm Exp}\left[{-\frac{\left(x+x_0+\frac{k_0}{\omega_{k_0}}t\right)^2}{4\sigma^2}}\right].
\eeq
We see that the kink's momentum has changed sign, now the part of the wave packet at $x<0$ is traveling in the opposite direction.  Also the position is $x=-x_t$, it has reflected from the kink.  However the entire wave packet has not reflected.  Here we have only calculated the contribution to the wave packet at $x<0$ and the amplitude has been damped by the factor $2B^*C$.  We interpret this factor as the reflection coefficient.

What happens on the right side of the kink?  One need only insert $\g_k(x)$ from Eq.~(\ref{gk}), at $x\gg 0$, into Eq.~(\ref{ft}) to obtain
%\bea
%\Phi(x,t)&=&2\sigma\sqrt{\pi}e^{-i\omega_{k_0}t}\pin{k} \left(
%B^*_k  e^{-i(k-k_0)\left(x_0+\frac{k_0}{\omega_{k_0}}t\right)}e^{-(k-k_0)^2\sigma^2}\right.\nonumber\\
%&&\left.+C^*_k e^{i(k+k_0)\left(x_0+\frac{k_0}{\omega_{k_0}}t\right)}e^{-(k+k_0)^2\sigma^2} 
%\right) \left(D_ke^{ikx}+E_k e^{-ikx}\right).
%\eea
\bea
\Phi(x,t)&=&e^{-i\omega_{k_0}t}\left(
\left(B^*_{k_0}D_{k_0}+C_{k_0}E^*_{k_0} \right){\rm Exp}\left[{-\frac{\left(-x+x_0+\frac{k_0}{\omega_{k_0}}t\right)^2}{4\sigma^2}}+ik_0 x\right]\right.\nonumber\\
&&
\left.+\left(B^*_{k_0}E_{k_0}+C_{k_0}D^*_{k_0}\right){\rm Exp}\left[{-\frac{\left(x+x_0+\frac{k_0}{\omega_{k_0}}t\right)^2}{4\sigma^2}}-ik_0 x\right]
\right).
\eea
The second line vanishes as a result of the constraint Eq.~(\ref{ort}).  The Gaussian on the first line has support when $x=x_t$.  

Since we are considering $x>0$, this only occurs when $x_t>0$ and so at $t\gg \omega_{k_0} |x_0|/k_0$, after the meson has interacted with the kink.  This is reasonable, the meson cannot get to the right of the kink before they have interacted.  At these late times, the position is $x_t$, implying that at this order the meson has continued to move past the kink at its initial velocity.  We interpret the $(B^*D+C E^*)$ factor as the transmission coefficient.

\section{Comparison with Quantum Mechanics}

The above calculation was equivalent to one in relativistic quantum mechanics.  This is because the kink frame of quantum field theory reduces the leading order problem of meson-kink scattering in the full quantum field theory to a simple quantum mechanical exercise.  Of course the above calculation has the advantage that it can be generalized to higher orders, revealing for example the backreaction of the meson on the kink, which is invisible in quantum mechanics where the kink is replaced by a potential well. 

But is the reflection coefficient really $2B^*C$?  Usually in quantum mechanics one does not use a normal mode basis for such a problem, as it complicates the derivation and obscures the physics.  It was only used here because of the simple connection to the kink-frame meson Fock space in Eq.~(\ref{fock}).

For simplicity, let us consider quantum mechanics with a symmetric potential.  This corresponds to a symmetric mass term in $H_2$, which arises from an antisymmetric kink in a symmetric potential in quantum field theory.  Beyond such cases, in general the quantum corrections to the vacuum energies on the two sides of the kink will disagree and so the kink will accelerate \cite{tstabile}, and will not correspond to a Hamiltonian eigenstate.  While such problems are invisible at the low order of perturbation theory considered here, and anyway do not prevent the application of our formalism, nonetheless they justify our interest in symmetric potentials.

In quantum mechanics, the kink at this order is replaced by a potential well and scattering is described by a symmetric and unitary $S$-matrix
\beq
S=\left(
\begin{tabular}{cc}
$t$&$r$\\
$r$&$t$\
\end{tabular}
\right)
\eeq
where $t$ and $r$ are the complex transmission and reflection coefficients such that, as a result of the unitarity of $S$,
\beq
|t|^2+|r|^2=1\hsp {\rm{Arg}}(r)={\rm{Arg}}(t)\pm \frac{\pi}{2}.
\eeq

Consider $k>0$.  Then $B_k$ and $E_k$ have the interpretations of particles incident on the well from the left and right respectively, while $C_k$ and $D_k$ correspond to particles moving away from the well on the left and right.  The $S$ matrix relates these
\beq
\left(\begin{tabular}{c}
$C_k$\\
$D_k$\\
\end{tabular}
\right)
=S\left(\begin{tabular}{c}
$E_k$\\
$B_k$\\
\end{tabular}
\right).
\eeq
The vectors $(C_k,D_k)$ and $(B_k,E_k)$ are orthogonal~(\ref{ortog}).  This fixes $(C_k,D_k)$ up to a phase $\phi$
\beq
S\left(\begin{tabular}{c}
$E_k$\\
$B_k$\\
\end{tabular}
\right)=\left(\begin{tabular}{c}
$C_k$\\
$D_k$\\
\end{tabular}
\right)
=e^{i\phi}\left(\begin{tabular}{c}
$-E_k$\\
$B_k$\\
\end{tabular}
\right).  \label{sapp}
\eeq
This linear set of equations is solved by
\beq
B_k=-\frac{t+e^{i\phi}}{r}E_k \hsp E_k= \frac{e^{i\phi}-t}{r}B_k. \label{be}
\eeq
Writing
\beq
t=|t|e^{i\theta}\hsp r=i\epsilon_1 |r|e^{i\theta}\hsp
\epsilon_1=\pm 1
\eeq
and combining the two equations (\ref{be}) one finds
\beq
E_k=\frac{e^{2i\phi}-|t|^2 e^{2i\theta}}{|r|^2 e^{2i\theta}} E_k
\eeq
implying
\beq
e^{i\phi}=\epsilon_2 e^{i\theta}\hsp
\epsilon_2=\pm 1
\eeq
and so
\beq
B_k=i\epsilon_1\epsilon_2\frac{1+\epsilon_2|t|}{|r|}E_k=
i\epsilon_1\epsilon_2\sqrt{\frac{1+\epsilon_2|t|}{1-\epsilon_2|t|}}E_k.
\eeq

The condition $|B_k|^2+|E_k|^2=1$ then leads to
\beq
|B_k|=\sqrt\frac{1+\epsilon_2|t|}{2}\hsp
|E_k|=\sqrt\frac{1-\epsilon_2|t|}{2}.
\eeq
Then, for some phase $\lambda$,
\beq
B_k=e^{i\lambda}\sqrt\frac{1+\epsilon_2|t|}{2}\hsp
E_k=-ie^{i\lambda}\epsilon_1\epsilon_2\sqrt\frac{1-\epsilon_2|t|}{2}.
\eeq

Now using the original equation (\ref{sapp}) we find
\beq
C_k=-e^{i\phi}E_k=ie^{i(\lambda+\phi)}\epsilon_1\epsilon_2\sqrt\frac{1-\epsilon_2|t|}{2}\hsp
D_k=e^{i\phi} B_k=e^{i(\lambda+\phi)}\sqrt\frac{1+\epsilon_2|t|}{2}.
\eeq
Therefore
\beq
2B^*_k C_k=2ie^{i\phi}\epsilon_1\epsilon_2\sqrt\frac{1+\epsilon_2|t|}{2}\sqrt\frac{1-\epsilon_2|t|}{2}=ie^{i\phi}\epsilon_1\epsilon_2|r|= r.
\eeq
Thus the quantum field theory coefficient for reflection, $2B^*C$, agrees with the usual result from quantum mechanics.   Similarly
\beq
B^*_kD_k+C_k E^*_k=e^{i\phi}\left(|B_k|^2-|E_k|^2\right)=e^{i\phi}\epsilon_2|t|=t
\eeq
and so the transmission coefficient also agrees with the usual result from quantum mechanics.

%Note however that the sign of $\epsilon_2$ flips if one shifts the sign of the reflected wave and simultaneously shifts the phase shift by $\pi$.  Therefore $2B^*C$ always agrees with the quantum mechanical reflection coefficient if one allows for a shift in the phase shift by $\pi$.

%The matrix $U$ remains unitary under a phase rotation of $B_k$ and $E_k$ so long as $C_k$ and $D_k$ rotate in the opposite direction.  Let us use such a rotation to make $B_k$ real and positive.  

\section{Remarks}

This calculation has been trivial.  It is a calculation that has been performed countless times in relativistic quantum mechanics over the past century, corresponding to the scattering of a particle through a symmetric barrier or well.  The interpretation of the calculation is somewhat novel, as the particle plays the role of the fundamental meson field $\phi(x)$ and the well is a leading approximation to a quantum kink.  The marvel of the kink frame is that it transforms a nonperturbative quantum field theory computation into an old exercise in quantum mechanics.  However it is likely expected that kink-meson scattering can, at leading order in the kink Hamiltonian, be treated in quantum mechanics.  

So what has been gained by repeating this old calculation?  We have extended the formalism of Refs.~\cite{mekink,me2loop} to include both time-evolution and also reflective potentials.  Reflective potentials are obviously interesting because they describe most kinks.  However, more to the point, we expect reflection at tree level even for kinks which are reflectionless in the free theory, and so this development is necessary in our opinion even for a perturbative treatment of reflectionless kinks such as those of the $\phi^4$ theory and the Sine-Gordon model.

More importantly, we have finally introduced time evolution into the formalism of linearized kink perturbation theory.  It has always used the Schrodinger picture, where operators are time-independent.  In addition, except for Ref.~\cite{chris}, it has restricted attention to Hamiltonian eigenstates, so that even the states do not evolve.  Even in Ref.~\cite{chris}, only time $t=0$ was considered, although instantaneous accelerations were computed.  The present note, on the other hand, describes for the first time how to apply this formalism to finite-time time evolution.  

The leading order state in quantum field theory has been found at each time $t$, even during the interaction itself although the asymptotic formula for $\g_k(x)$ cannot be applied there.  The analytic formula for $\g_k(x)$ are anyway known in the Sine-Gordon and $\phi^4$ models, and very good numerical approximations have been found in many other models. Therefore Eqs.(\ref{alf}) and (\ref{ft}) indeed give the full state, at leading order in the semiclassical expansion, at any time during the evolution.  This is progress with respect to Euclidean time methods, which are generally applied to the calculation of the $S$-matrix and so are essentially restricted to infinite-time evolution.  These leading order expressions for the states at each time, as well as the general formalism for kink-frame time evolution introduced here, of course are necessary to compute the perturbative corrections, which include the new physics that we intend to investigate in the near future.

\section* {Acknowledgement}

\noindent
JE is supported by NSFC MianShang grants 11875296 and 11675223. HL acknowledges the support from CAS-DAAD Joint Fellowship Programme for Doctoral students of UCAS.


\begin{thebibliography}{99}

\bibitem{csw} D.~K. Campbell, J.~F. Schonfeld and C.~A. Wingate,
``Resonance structure in kink-antikink interactions in $\phi^4$ theory,"
Physica \textbf{D9} (1983) 1.

\bibitem{doreynoshape}
P.~Dorey, K.~Mersh, T.~Romanczukiewicz and Y.~Shnir,
``Kink-antikink collisions in the $\phi^6$ model,''
Phys. Rev. Lett. \textbf{107} (2011), 091602
doi:10.1103/PhysRevLett.107.091602
[arXiv:1101.5951 [hep-th]].

\bibitem{doreyinterp}
P.~Dorey, A.~Gorina, I.~Perapechka, T.~Roma\'nczukiewicz and Y.~Shnir,
``Resonance structures in kink-antikink collisions in a deformed sine-Gordon model,''
JHEP \textbf{09} (2021), 145
doi:10.1007/JHEP09(2021)145
[arXiv:2106.09560 [hep-th]].

\bibitem{tomcc}
C.~Adam, P.~Dorey, A.~Garcia Martin-Caro, M.~Huidobro, K.~Oles, T.~Romanczukiewicz, Y.~Shnir and A.~Wereszczynski,
``Multikink scattering in the $\phi^6$ model revisited,''
[arXiv:2209.08849 [hep-th]].

\bibitem{faddeev77}
L.~D.~Faddeev and V.~E.~Korepin,
``Quantum Theory of Solitons: Preliminary Version,''
Phys. Rept. \textbf{42} (1978), 1-87
doi:10.1016/0370-1573(78)90058-3

\bibitem{lowe79}
M.~Lowe,
``BOSON - SOLITON SCATTERING IN THE SINE-GORDON MODEL,''
Nucl. Phys. B \textbf{159} (1979), 349-362
doi:10.1016/0550-3213(79)90339-0

\bibitem{parm87}
J.~A.~Parmentola and I.~Zahed,
``MESON - SOLITON SCATTERING WITH SOLITON RECOIL,''
Print-87-0301 (STONY BROOK).

\bibitem{swanson88}
M.~S.~Swanson,
``SOLITON-PARTICLE SCATTERING AND BERRY'S PHASE,''
Phys. Rev. D \textbf{38} (1988), 3122-3127
doi:10.1103/PhysRevD.38.3122

\bibitem{uehara91}
M.~Uehara, A.~Hayashi and S.~Saito,
``Meson - soliton scattering with full recoil in standard collective coordinate quantization,''
Nucl. Phys. A \textbf{534} (1991), 680-696
doi:10.1016/0375-9474(91)90466-J

\bibitem{abdel11}
A.~M.~H.~H.~Abdelhady and H.~Weigel,
``Wave-Packet Scattering off the Kink-Solution,''
Int. J. Mod. Phys. A \textbf{26} (2011), 3625-3640
doi:10.1142/S0217751X11054012
[arXiv:1106.3497 [nlin.PS]].

\bibitem{tomrad1}
T.~Romanczukiewicz,
``Interaction between kink and radiation in phi**4 model,''
Acta Phys. Polon. B \textbf{35} (2004), 523-540
[arXiv:hep-th/0303058 [hep-th]].

\bibitem{tomrad2}
T.~Romanczukiewicz,
``Interaction between topological defects and radiation,''
Acta Phys. Polon. B \textbf{36} (2005), 3877-3887

\bibitem{tomrad3}
P.~Forgacs, A.~Lukacs and T.~Romanczukiewicz,
``Negative radiation pressure exerted on kinks,''
Phys. Rev. D \textbf{77} (2008), 125012
doi:10.1103/PhysRevD.77.125012
[arXiv:0802.0080 [hep-th]].

\bibitem{dhn2}
  R.~F.~Dashen, B.~Hasslacher and A.~Neveu,
  ``Nonperturbative Methods and Extended Hadron Models in Field Theory 2. Two-Dimensional Models and Extended Hadrons,''
  Phys.\ Rev.\ D {\bf 10} (1974) 4130.
 doi:10.1103/PhysRevD.10.4130


\bibitem{gjscc}
J.~L.~Gervais, A.~Jevicki and B.~Sakita,
``Collective Coordinate Method for Quantization of Extended Systems,''
Phys. Rept. \textbf{23} (1976), 281-293
doi:10.1016/0370-1573(76)90049-1

\bibitem{gj76}
J.~L.~Gervais and A.~Jevicki,
``Point Canonical Transformations in Path Integral,''
Nucl. Phys. B \textbf{110} (1976), 93-112
doi:10.1016/0550-3213(76)90422-3

\bibitem{mekink}
J.~Evslin,
``Manifestly Finite Derivation of the Quantum Kink Mass,''
JHEP \textbf{11} (2019), 161
doi:10.1007/JHEP11(2019)161
[arXiv:1908.06710 [hep-th]].

\bibitem{me2loop}
J.~Evslin and H.~Guo,
``Two-Loop Scalar Kinks,''
Phys. Rev. D \textbf{103} (2021) no.12, 125011
doi:10.1103/PhysRevD.103.125011
[arXiv:2012.04912 [hep-th]].

\bibitem{chris}
J.~Evslin, C.~Halcrow, T.~Romanczukiewicz and A.~Wereszczynski,
``Spectral walls at one loop,''
Phys. Rev. D \textbf{105} (2022) no.12, 125002
doi:10.1103/PhysRevD.105.125002
[arXiv:2202.08249 [hep-th]].

\bibitem{weigelstab}
H.~Weigel,
``Quantum Instabilities of Solitons,''
AIP Conf. Proc. \textbf{2116} (2019) no.1, 170002
doi:10.1063/1.5114153
[arXiv:1907.10942 [hep-th]].

\bibitem{rebhan}
  A.~Rebhan and P.~van Nieuwenhuizen,
  ``No saturation of the quantum Bogomolnyi bound by two-dimensional supersymmetric solitons,''
  Nucl.\ Phys.\ B {\bf 508} (1997) 449
  doi:10.1016/S0550-3213(97)00625-1, 10.1016/S0550-3213(97)80021-1
 [hep-th/9707163].

\bibitem{rob}
Rob Pisarski, Private communication.

\bibitem{cahill76}
K.~E.~Cahill, A.~Comtet and R.~J.~Glauber,
``Mass Formulas for Static Solitons,''
Phys. Lett. B \textbf{64} (1976), 283-285
doi:10.1016/0370-2693(76)90202-1

\bibitem{tstabile}
T.~Roma\'nczukiewicz,
``Could the primordial radiation be responsible for vanishing of topological impuritys?,''
Phys. Lett. B \textbf{773} (2017), 295-299
doi:10.1016/j.physletb.2017.08.045
[arXiv:1706.05192 [hep-th]].

\end{thebibliography}

\begin{thebibliography}{99}

\bibitem{mm}
N.~S.~Manton and H.~Merabet,
``$\phi^4$ kinks: Gradient flow and dynamics,''
Nonlinearity \textbf{10} (1997), 3
doi:10.1088/0951-7715/10/1/002
[arXiv:hep-th/9605038 [hep-th]].

\bibitem{chris}
J.~Evslin, C.~Halcrow, T.~Romanczukiewicz and A.~Wereszczynski,
``Spectral walls at one loop,''
Phys. Rev. D \textbf{105} (2022) no.12, 125002
doi:10.1103/PhysRevD.105.125002
[arXiv:2202.08249 [hep-th]].

\bibitem{phi42loop}
J.~Evslin,
``\ensuremath{\phi^4} kink mass at two loops,''
Phys. Rev. D \textbf{104} (2021) no.8, 085013
doi:10.1103/PhysRevD.104.085013
[arXiv:2104.07991 [hep-th]].


\bibitem{decad0}
M.~Lévy,  ``On the description of unstable particles in quantum field theory,'' Nuovo Cim \textbf{13} (1959) 115-143
doi.org/10.1007/BF02726390

\bibitem{decad1}
M.~Lévy,  ``On the validity of the exponential law for the decay of an unstable particle,'' Nuovo Cim \textbf{14} (1959) 612–624
doi.org/10.1007/BF02726390

\bibitem{decad2}
M.~Ram,
``Mass and total decay rate of unstable particles in quantum field theory,''
Int. J. Theor. Phys. \textbf{3} (1970), 153-164
doi:10.1007/BF02412757

\bibitem{me2loop}
J.~Evslin and H.~Guo,
``Two-Loop Scalar Kinks,''
Phys. Rev. D \textbf{103} (2021) no.12, 125011
doi:10.1103/PhysRevD.103.125011
[arXiv:2012.04912 [hep-th]].

\bibitem{rebhan}
  A.~Rebhan and P.~van Nieuwenhuizen,
  ``No saturation of the quantum Bogomolnyi bound by two-dimensional supersymmetric solitons,''
  Nucl.\ Phys.\ B {\bf 508} (1997) 449
  doi:10.1016/S0550-3213(97)00625-1, 10.1016/S0550-3213(97)80021-1
 [hep-th/9707163].

\end{thebibliography}

\begin{thebibliography}{99}

\bibitem{weigelstab}
H.~Weigel,
``Quantum Instabilities of Solitons,''
AIP Conf. Proc. \textbf{2116} (2019) no.1, 170002
doi:10.1063/1.5114153
[arXiv:1907.10942 [hep-th]].


\bibitem{delfino}
  G.~Delfino, W.~Selke and A.~Squarcini,
  ``Vortex mass in the three-dimensional $O(2)$ scalar theory,''
  Phys.\ Rev.\ Lett.\  {\bf 122} (2019) no.5,  050602
  doi:10.1103/PhysRevLett.122.050602
  [arXiv:1808.09276 [cond-mat.stat-mech]].

\bibitem{davies}
  D.~Davies,
  ``Quantum Solitons in any Dimension: Derrick's Theorem v. AQFT,''
  arXiv:1907.10616 [hep-th].

\bibitem{dhn2}
  R.~F.~Dashen, B.~Hasslacher and A.~Neveu,
  ``Nonperturbative Methods and Extended Hadron Models in Field Theory 2. Two-Dimensional Models and Extended Hadrons,''
  Phys.\ Rev.\ D {\bf 10} (1974) 4130.
 doi:10.1103/PhysRevD.10.4130

\bibitem{rajaraman75}
R.~Rajaraman,
``Some Nonperturbative Semiclassical Methods in Quantum Field Theory: A Pedagogical Review,''
Phys. Rept. \textbf{21} (1975), 227-313
doi:10.1016/0370-1573(75)90016-2

\bibitem{goldhaber04}
A.~S.~Goldhaber, A.~Rebhan, P.~van Nieuwenhuizen and R.~Wimmer,
``Quantum corrections to mass and central charge of supersymmetric solitons,''
Phys. Rept. \textbf{398} (2004), 179-219
doi:10.1016/j.physrep.2004.05.001
[arXiv:hep-th/0401152 [hep-th]].

\bibitem{weigrev}
N.~Graham and H.~Weigel,
``Quantum Corrections to Soliton Energies,''
[arXiv:2201.12131 [hep-th]].

\bibitem{gjscc}
J.~L.~Gervais, A.~Jevicki and B.~Sakita,
``Collective Coordinate Method for Quantization of Extended Systems,''
Phys. Rept. \textbf{23} (1976), 281-293
doi:10.1016/0370-1573(76)90049-1

\bibitem{gj76}
J.~L.~Gervais and A.~Jevicki,
``Point Canonical Transformations in Path Integral,''
Nucl. Phys. B \textbf{110} (1976), 93-112
doi:10.1016/0550-3213(76)90422-3

\bibitem{vega}
H.~J.~de Vega,
``Two-Loop Quantum Corrections to the Soliton Mass in Two-Dimensional Scalar Field Theories,''
Nucl. Phys. B \textbf{115} (1976), 411-428
doi:10.1016/0550-3213(76)90497-1

%\cite{Verwaest:1977tw}
\bibitem{verwaest}
J.~Verwaest,
``Higher Order Correction to the Sine-Gordon Soliton Mass,''
Nucl. Phys. B \textbf{123} (1977), 100-108
doi:10.1016/0550-3213(77)90343-1

\bibitem{shif99}
M.~A.~Shifman, A.~I.~Vainshtein and M.~B.~Voloshin,
``Anomaly and quantum corrections to solitons in two-dimensional theories with minimal supersymmetry,''
Phys. Rev. D \textbf{59} (1999), 045016
doi:10.1103/PhysRevD.59.045016
[arXiv:hep-th/9810068 [hep-th]].

\bibitem{me2loop}
J.~Evslin and H.~Guo,
``Two-Loop Scalar Kinks,''
Phys. Rev. D \textbf{103} (2021) no.12, 125011
doi:10.1103/PhysRevD.103.125011
[arXiv:2012.04912 [hep-th]].

\bibitem{mephi4}
J.~Evslin,
``\ensuremath{\phi}${}^4$ kink mass at two loops,''
Phys. Rev. D \textbf{104} (2021) no.8, 085013
doi:10.1103/PhysRevD.104.085013
[arXiv:2104.07991 [hep-th]].

\bibitem{me2loopex}
J.~Evslin and H.~Guo,
``Excited Kinks as Quantum States,''
Eur. Phys. J. C \textbf{81} (2021) no.10, 936
doi:10.1140/epjc/s10052-021-09739-9
[arXiv:2104.03612 [hep-th]].

\bibitem{chris}
J.~Evslin, C.~Halcrow, T.~Romanczukiewicz and A.~Wereszczynski,
``Spectral Walls at One Loop,''
[arXiv:2202.08249 [hep-th]].

\bibitem{meimpulso}
J.~Evslin,
``Moving Kinks and Their Wave Packets,''
[arXiv:2202.04905 [hep-th]].

\bibitem{skyrme}
T.~H.~R.~Skyrme,
``A Nonlinear field theory,''
Proc. Roy. Soc. Lond. A \textbf{260} (1961), 127-138
doi:10.1098/rspa.1961.0018

\bibitem{wittenskyrme}
E.~Witten,
``Current Algebra, Baryons, and Quark Confinement,''
Nucl. Phys. B \textbf{223} (1983), 433-444
doi:10.1016/0550-3213(83)90064-0

\bibitem{smorg}
S.~B.~Gudnason and C.~Halcrow,
``A Sm\"orgasbord of Skyrmions,''
[arXiv:2202.01792 [hep-th]].

\bibitem{faddeev77}
L.~D.~Faddeev and V.~E.~Korepin,
``Quantum Theory of Solitons: Preliminary Version,''
Phys. Rept. \textbf{42} (1978), 1-87
doi:10.1016/0370-1573(78)90058-3

\bibitem{lowe79}
M.~Lowe,
``BOSON - SOLITON SCATTERING IN THE SINE-GORDON MODEL,''
Nucl. Phys. B \textbf{159} (1979), 349-362
doi:10.1016/0550-3213(79)90339-0

\bibitem{parm87}
J.~A.~Parmentola and I.~Zahed,
``MESON - SOLITON SCATTERING WITH SOLITON RECOIL,''
Print-87-0301 (STONY BROOK).

\bibitem{swanson88}
M.~S.~Swanson,
``SOLITON-PARTICLE SCATTERING AND BERRY'S PHASE,''
Phys. Rev. D \textbf{38} (1988), 3122-3127
doi:10.1103/PhysRevD.38.3122

\bibitem{uehara91}
M.~Uehara, A.~Hayashi and S.~Saito,
``Meson - soliton scattering with full recoil in standard collective coordinate quantization,''
Nucl. Phys. A \textbf{534} (1991), 680-696
doi:10.1016/0375-9474(91)90466-J

\bibitem{abdel11}
A.~M.~H.~H.~Abdelhady and H.~Weigel,
``Wave-Packet Scattering off the Kink-Solution,''
Int. J. Mod. Phys. A \textbf{26} (2011), 3625-3640
doi:10.1142/S0217751X11054012
[arXiv:1106.3497 [nlin.PS]].

\bibitem{andyff2}
I.~V.~Melnikov, C.~Papageorgakis and A.~B.~Royston,
``Forced Soliton Equation and Semiclassical Soliton Form Factors,''
Phys. Rev. Lett. \textbf{125} (2020) no.23, 231601
doi:10.1103/PhysRevLett.125.231601
[arXiv:2010.10381 [hep-th]].

\bibitem{loginov22}
A.~Y.~Loginov,
``Scattering of fermionic isodoublets on the sine-Gordon kink,''
[arXiv:2202.13086 [hep-th]].


\bibitem{gj75}
J.~Goldstone and R.~Jackiw,
``Quantization of Nonlinear Waves,''
Phys. Rev. D \textbf{11} (1975), 1486-1498
doi:10.1103/PhysRevD.11.1486


\bibitem{weisz77}
P.~H.~Weisz,
``Exact Quantum Sine-Gordon Soliton Form-Factors,''
Phys. Lett. B \textbf{67} (1977), 179-182
doi:10.1016/0370-2693(77)90097-1

\bibitem{bab01}
H.~Babujian and M.~Karowski,
``Exact form-factors in integrable quantum field theories: The sine-Gordon model. 2.,''
Nucl. Phys. B \textbf{620} (2002), 407-455
doi:10.1016/S0550-3213(01)00551-X
[arXiv:hep-th/0105178 [hep-th]].


\bibitem{rebhan}
  A.~Rebhan and P.~van Nieuwenhuizen,
  ``No saturation of the quantum Bogomolnyi bound by two-dimensional supersymmetric solitons,''
  Nucl.\ Phys.\ B {\bf 508} (1997) 449
  doi:10.1016/S0550-3213(97)00625-1, 10.1016/S0550-3213(97)80021-1
 [hep-th/9707163].

\bibitem{tstabile}
T.~Roma\'nczukiewicz,
``Could the primordial radiation be responsible for vanishing of topological impuritys?,''
Phys. Lett. B \textbf{773} (2017), 295-299
doi:10.1016/j.physletb.2017.08.045
[arXiv:1706.05192 [hep-th]].

\bibitem{wstabile}
H.~Weigel,
``Quantum Instabilities of Solitons,''
AIP Conf. Proc. \textbf{2116} (2019) no.1, 170002
doi:10.1063/1.5114153
[arXiv:1907.10942 [hep-th]].

\bibitem{mekink}
J.~Evslin,
``Manifestly Finite Derivation of the Quantum Kink Mass,''
JHEP \textbf{11} (2019), 161
doi:10.1007/JHEP11(2019)161
[arXiv:1908.06710 [hep-th]].

\bibitem{cahill76}
K.~E.~Cahill, A.~Comtet and R.~J.~Glauber,
``Mass Formulas for Static Solitons,''
Phys. Lett. B \textbf{64} (1976), 283-285
doi:10.1016/0370-2693(76)90202-1

\bibitem{mewick}
J.~Evslin,
``Normal ordering normal modes,''
Eur. Phys. J. C \textbf{81} (2021) no.1, 92
doi:10.1140/epjc/s10052-021-08890-7
[arXiv:2007.05741 [hep-th]].

\bibitem{gervaisff75}
J.~L.~Gervais, A.~Jevicki and B.~Sakita,
``Perturbation Expansion Around Extended Particle States in Quantum Field Theory. 1.,''
Phys. Rev. D \textbf{12} (1975), 1038
doi:10.1103/PhysRevD.12.1038

\bibitem{megirini}
J.~Evslin and H.~Guo,
``Removing tadpoles in a soliton sector,''
JHEP \textbf{11} (2021), 128
doi:10.1007/JHEP11(2021)128
[arXiv:2110.00234 [hep-th]].

\bibitem{smirnov92}
F.~A.~Smirnov,
``Form-factors in completely integrable models of quantum field theory,''
Adv. Ser. Math. Phys. \textbf{14} (1992), 1-208

\bibitem{hayashi92}
A.~Hayashi, S.~Saito and M.~Uehara,
``Pion - nucleon scattering in the soliton model,''
Prog. Theor. Phys. Suppl. \textbf{109} (1992), 45-72
doi:10.1143/PTPS.109.45

\bibitem{zam77}
A.~B.~Zamolodchikov,
``Exact Two Particle s Matrix of Quantum Sine-Gordon Solitons,''
Pisma Zh. Eksp. Teor. Fiz. \textbf{25} (1977), 499-502
doi:10.1007/BF01626520

\end{thebibliography}

\begin{thebibliography}{99}

\end{thebibliography}

\begin{thebibliography}{99}

\bibitem{me2loop}
J.~Evslin and H.~Guo,
``Two-Loop Scalar Kinks,''
Phys. Rev. D \textbf{103} (2021) no.12, 125011
doi:10.1103/PhysRevD.103.125011
[arXiv:2012.04912 [hep-th]].


\bibitem{weigrev}
N.~Graham and H.~Weigel,
``Quantum Corrections to Soliton Energies,''
[arXiv:2201.12131 [hep-th]].

\bibitem{mestato}
J.~Evslin,
``The Ground State of the Sine-Gordon Soliton,''
JHEP \textbf{07} (2020), 099
doi:10.1007/JHEP07(2020)099
[arXiv:2003.11384 [hep-th]].

\bibitem{memassa}
J.~Evslin,
``Well-defined quantum soliton masses without supersymmetry,''
Phys. Rev. D \textbf{101} (2020) no.6, 065005
doi:10.1103/PhysRevD.101.065005
[arXiv:2002.12523 [hep-th]].

\bibitem{chris}
J.~Evslin, Chris Halcrow, Tomasz Romanczukiewicz and Andrzej Wereszczynski,
``Spectral Walls at One Loop,''
in preparation.

\bibitem{impure19}
C.~Adam, T.~Romanczukiewicz and A.~Wereszczynski,
``The $\phi^4$ model with the BPS preserving defect,''
JHEP \textbf{03} (2019), 131
doi:10.1007/JHEP03(2019)131
[arXiv:1812.04007 [hep-th]].

\bibitem{muri}
C.~Adam, K.~Oles, T.~Romanczukiewicz and A.~Wereszczynski,
``Spectral Walls in Soliton Collisions'',
Phys. Rev. Lett. \textbf{122} (2019) no.24, 241601
doi:10.1103/PhysRevLett.122.241601
[arXiv:1903.12100].

\bibitem{moduli}
N.~S.~Manton, K.~Ole\'s, T.~Roma\'nczukiewicz and A.~Wereszczy\'nski,
``Kink moduli spaces: Collective coordinates reconsidered,''
Phys. Rev. D \textbf{103} (2021) no.2, 025024
doi:10.1103/PhysRevD.103.025024
[arXiv:2008.01026 [hep-th]].

%\cite{Skyrme:1961vq}
\bibitem{skyrme}
T.~H.~R.~Skyrme,
``A Nonlinear field theory,''
Proc. Roy. Soc. Lond. A \textbf{260} (1961), 127-138
doi:10.1098/rspa.1961.0018

\bibitem{wittenskyrme}
E.~Witten,
``Current Algebra, Baryons, and Quark Confinement,''
Nucl. Phys. B \textbf{223} (1983), 433-444
doi:10.1016/0550-3213(83)90064-0

\bibitem{smorg}
S.~B.~Gudnason and C.~Halcrow,
``A Sm\"orgasbord of Skyrmions,''
[arXiv:2202.01792 [hep-th]].

\bibitem{medark}
J.~Evslin and S.~B.~Gudnason,
``Dwarf Galaxy Sized Monopoles as Dark Matter?,''
[arXiv:1202.0560 [astro-ph.CO]].

\bibitem{lorodark}
H.~Y.~Schive, T.~Chiueh and T.~Broadhurst,
``Cosmic Structure as the Quantum Interference of a Coherent Dark Wave,''
Nature Phys. \textbf{10} (2014), 496-499
doi:10.1038/nphys2996
[arXiv:1406.6586 [astro-ph.GA]].

\bibitem{abrik}
A. A. Abrikosov, 
``The magnetic properties of superconducting alloys,''
Journal of Physics and Chemistry of Solids. 2  (1957) 199–208 
doi:10.1016/0022-3697(57)90083-5.

\bibitem{optix}
J. E. Bjorkholm and A. A. Ashkin, 
``Self-Focusing and Self-Trapping of Light in Sodium Vapor,''
Phys. Rev. Lett. 32 (1974) 129-132
doi:10.1103/PhysRevLett.32.129.

\bibitem{dhn2}
  R.~F.~Dashen, B.~Hasslacher and A.~Neveu,
  ``Nonperturbative Methods and Extended Hadron Models in Field Theory 2. Two-Dimensional Models and Extended Hadrons,''
  Phys.\ Rev.\ D {\bf 10} (1974) 4130.
 doi:10.1103/PhysRevD.10.4130

\bibitem{rebhan}
  A.~Rebhan and P.~van Nieuwenhuizen,
  ``No saturation of the quantum Bogomolnyi bound by two-dimensional supersymmetric solitons,''
  Nucl.\ Phys.\ B {\bf 508} (1997) 449
  doi:10.1016/S0550-3213(97)00625-1, 10.1016/S0550-3213(97)80021-1
 [hep-th/9707163].

\bibitem{mekink}
J.~Evslin,
``Manifestly Finite Derivation of the Quantum Kink Mass,''
JHEP \textbf{11} (2019), 161
doi:10.1007/JHEP11(2019)161
[arXiv:1908.06710 [hep-th]].


\bibitem{cahill76}
K.~E.~Cahill, A.~Comtet and R.~J.~Glauber,
``Mass Formulas for Static Solitons,''
Phys. Lett. B \textbf{64} (1976), 283-285
doi:10.1016/0370-2693(76)90202-1

\bibitem{gjscc}
J.~L.~Gervais, A.~Jevicki and B.~Sakita,
``Collective Coordinate Method for Quantization of Extended Systems,''
Phys. Rept. \textbf{23} (1976), 281-293
doi:10.1016/0370-1573(76)90049-1


\bibitem{tstabile}
T.~Roma\'nczukiewicz,
``Could the primordial radiation be responsible for vanishing of topological impuritys?,''
Phys. Lett. B \textbf{773} (2017), 295-299
doi:10.1016/j.physletb.2017.08.045
[arXiv:1706.05192 [hep-th]].

\bibitem{wstabile}
H.~Weigel,
``Quantum Instabilities of Solitons,''
AIP Conf. Proc. \textbf{2116} (2019) no.1, 170002
doi:10.1063/1.5114153
[arXiv:1907.10942 [hep-th]].

\bibitem{wpol}
H.~Weigel,
``Vacuum Polarization Energy for General Backgrounds in One Space Dimension,''
Phys. Lett. B \textbf{766} (2017), 65-70
doi:10.1016/j.physletb.2016.12.055
[arXiv:1612.08641 [hep-th]].

\bibitem{wentzel}
G. Wentzel, 
``Zur Paartheorie der Kernkr\"afte,''
Helv. Phys. Acta 15 (1942) 111.



%\bibitem{meplb}
%J.~Evslin,
%``The two-loop \ensuremath{\phi}4 kink mass,''
%Phys. Lett. B \textbf{822} (2021), 136628
%doi:10.1016/j.physletb.2021.136628
%[arXiv:2109.05852 [hep-th]].

\bibitem{mephi4}
J.~Evslin,
``\ensuremath{\phi}${}^4$ kink mass at two loops,''
Phys. Rev. D \textbf{104} (2021) no.8, 085013
doi:10.1103/PhysRevD.104.085013
[arXiv:2104.07991 [hep-th]].

\bibitem{mewick}
J.~Evslin,
``Normal Ordering Normal Modes,''
Eur. Phys. J. C \textbf{81} (2021) no.1, 92
doi:10.1140/epjc/s10052-021-08890-7
[arXiv:2007.05741 [hep-th]].



\bibitem{weisz77}
P.~H.~Weisz,
``Exact Quantum Sine-Gordon Soliton Form-Factors,''
Phys. Lett. B \textbf{67} (1977), 179-182
doi:10.1016/0370-2693(77)90097-1

\bibitem{smirnov92}
F.~A.~Smirnov,
``Form-factors in completely integrable models of quantum field theory,''
Adv. Ser. Math. Phys. \textbf{14} (1992), 1-208

\bibitem{bab01}
H.~Babujian and M.~Karowski,
``Exact form-factors in integrable quantum field theories: The sine-Gordon model. 2.,''
Nucl. Phys. B \textbf{620} (2002), 407-455
doi:10.1016/S0550-3213(01)00551-X
[arXiv:hep-th/0105178 [hep-th]].

\bibitem{andyff2}
I.~V.~Melnikov, C.~Papageorgakis and A.~B.~Royston,
``Forced Soliton Equation and Semiclassical Soliton Form Factors,''
Phys. Rev. Lett. \textbf{125} (2020) no.23, 231601
doi:10.1103/PhysRevLett.125.231601
[arXiv:2010.10381 [hep-th]].



\bibitem{kovtun}
A.~Kovtun,
``Analytical computation of quantum corrections to non-topological soliton (bright soliton) within the saddle-point approximation,''
[arXiv:2110.05222 [hep-th]].

\bibitem{yuan21}
Y.~Zhong, F.~Y.~Li and X.~D.~Liu,
``K-field kinks in two-dimensional dilaton gravity,''
Phys. Lett. B \textbf{822} (2021), 136716
doi:10.1016/j.physletb.2021.136716
[arXiv:2108.10166 [hep-th]].

\bibitem{yuannor1}
Y.~Zhong,
``Normal modes for two-dimensional gravitating kinks,''
[arXiv:2112.08683 [hep-th]].

\bibitem{yuannor2}
J.~Feng and Y.~Zhong,
``Scalar perturbation of gravitating double-kink solutions,''
[arXiv:2202.02946 [hep-th]].

\end{thebibliography}

\begin{thebibliography}{99}
	

\bibitem{dhn2}
  R.~F.~Dashen, B.~Hasslacher and A.~Neveu,
  ``Nonperturbative Methods and Extended Hadron Models in Field Theory 2. Two-Dimensional Models and Extended Hadrons,''
  Phys.\ Rev.\ D {\bf 10} (1974) 4130.
 doi:10.1103/PhysRevD.10.4130

\bibitem{vega}
H.~J.~de Vega,
``Two-Loop Quantum Corrections to the Soliton Mass in Two-Dimensional Scalar Field Theories,''
Nucl. Phys. B \textbf{115} (1976), 411-428
doi:10.1016/0550-3213(76)90497-1

%\cite{Verwaest:1977tw}
\bibitem{verwaest}
J.~Verwaest,
``Higher Order Correction to the Sine-Gordon Soliton Mass,''
Nucl. Phys. B \textbf{123} (1977), 100-108
doi:10.1016/0550-3213(77)90343-1

\bibitem{menormal}
J.~Evslin and H.~Guo,
``Excited Kinks as Quantum States,''
[arXiv:2104.03612 [hep-th]].

\bibitem{meshape}
J.~Evslin,
``Evidence for the unbinding of the $\phi^{4}$ kink's shape mode,''
JHEP \textbf{09} (2021), 009
doi:10.1007/JHEP09(2021)009
[arXiv:2104.14387 [hep-th]].

\bibitem{mekink}
J.~Evslin,
``Manifestly Finite Derivation of the Quantum Kink Mass,''
JHEP \textbf{11} (2019), 161
doi:10.1007/JHEP11(2019)161
[arXiv:1908.06710 [hep-th]].

\bibitem{mephi4massa}
J.~Evslin,
``The two-loop $\phi^4$ kink mass,''
Phys. Lett. B \textbf{822} (2021), 136628
doi:10.1016/j.physletb.2021.136628
[arXiv:2109.05852 [hep-th]].

\bibitem{tamasstab}
T.~Roma\'nczukiewicz,
``Could the primordial radiation be responsible for vanishing of topological defects?,''
Phys. Lett. B \textbf{773} (2017), 295-299
doi:10.1016/j.physletb.2017.08.045
[arXiv:1706.05192 [hep-th]].

\bibitem{weigelstab}
H.~Weigel,
``Quantum Instabilities of Solitons,''
AIP Conf. Proc. \textbf{2116} (2019) no.1, 170002
doi:10.1063/1.5114153
[arXiv:1907.10942 [hep-th]].


\bibitem{cahill76}
K.~E.~Cahill, A.~Comtet and R.~J.~Glauber,
``Mass Formulas for Static Solitons,''
Phys. Lett. B \textbf{64} (1976), 283-285
doi:10.1016/0370-2693(76)90202-1

\bibitem{memassa}
J.~Evslin,
``Well-defined quantum soliton masses without supersymmetry,''
Phys. Rev. D \textbf{101} (2020) no.6, 065005
doi:10.1103/PhysRevD.101.065005
[arXiv:2002.12523 [hep-th]].

\bibitem{metwoloop}
J.~Evslin and H.~Guo,
``Alternative to collective coordinates,''
Phys. Rev. D \textbf{103} (2021) no.4, L041701
doi:10.1103/PhysRevD.103.L041701
[arXiv:2101.08028 [hep-th]].

\bibitem{mewick}
J.~Evslin,
``Normal ordering normal modes,''
Eur. Phys. J. C \textbf{81} (2021) no.1, 92
doi:10.1140/epjc/s10052-021-08890-7
[arXiv:2007.05741 [hep-th]].

\bibitem{weiszff}
P.~H.~Weisz,
``Exact Quantum Sine-Gordon Soliton Form-Factors,''
Phys. Lett. B \textbf{67} (1977), 179-182
doi:10.1016/0370-2693(77)90097-1

\bibitem{karowskiff}
M.~Karowski and P.~Weisz,
``Exact Form-Factors in (1+1)-Dimensional Field Theoretic Models with Soliton Behavior,''
Nucl. Phys. B \textbf{139} (1978), 455-476
doi:10.1016/0550-3213(78)90362-0

\bibitem{babujianff}
H.~Babujian and M.~Karowski,
``Exact form-factors in integrable quantum field theories: The sine-Gordon model. 2.,''
Nucl. Phys. B \textbf{620} (2002), 407-455
doi:10.1016/S0550-3213(01)00551-X
[arXiv:hep-th/0105178 [hep-th]].

\bibitem{andyff1}
I.~V.~Melnikov, C.~Papageorgakis and A.~B.~Royston,
``Accelerating solitons,''
Phys. Rev. D \textbf{102} (2020) no.12, 125002
doi:10.1103/PhysRevD.102.125002
[arXiv:2007.11028 [hep-th]].

\bibitem{andyff2}
I.~V.~Melnikov, C.~Papageorgakis and A.~B.~Royston,
``Forced Soliton Equation and Semiclassical Soliton Form Factors,''
Phys. Rev. Lett. \textbf{125} (2020) no.23, 231601
doi:10.1103/PhysRevLett.125.231601
[arXiv:2010.10381 [hep-th]].

\bibitem{muri1}
C.~Adam, K.~Oles, T.~Romanczukiewicz and A.~Wereszczynski,
``Spectral Walls in Soliton Collisions,''
Phys. Rev. Lett. \textbf{122} (2019) no.24, 241601
doi:10.1103/PhysRevLett.122.241601
[arXiv:1903.12100 [hep-th]].

\bibitem{muri2}
C.~Adam, K.~Oles, T.~Romanczukiewicz, A.~Wereszczynski and W.~J.~Zakrzewski,
``Spectral walls in multifield kink dynamics,''
JHEP \textbf{08} (2021), 147
doi:10.1007/JHEP08(2021)147
[arXiv:2105.14771 [hep-th]].

\bibitem{gj75}
J.~Goldstone and R.~Jackiw,
``Quantization of Nonlinear Waves,''
Phys. Rev. D \textbf{11} (1975), 1486-1498
doi:10.1103/PhysRevD.11.1486

\bibitem{gjs}
J.~L.~Gervais, A.~Jevicki and B.~Sakita,
``Perturbation Expansion Around Extended Particle States in Quantum Field Theory. 1.,''
Phys. Rev. D \textbf{12} (1975), 1038
doi:10.1103/PhysRevD.12.1038

\end{thebibliography}

\begin{thebibliography}{99}
	
\bibitem{coleman}
S.Coleman,
``Quantum Field Theory Lecture of Sidney Coleman''(2019).World Scientific

\bibitem{me2loopex}
J.~Evslin and H.~Guo,
``Excited Kinks as Quantum States,''
[arXiv:2104.03612 [hep-th]].

\bibitem{salam}
Abdus.Salam
``some Speccuations on the New Resonances,''
Rev. Mod. Phys. \textbf{33}, 426-430 (1961)
doi:10.1103/RevModPhys.33.426

\bibitem{sakurai}
J.~J.~Sakurai,
``Dynamical Approach to the Unitary-Symmetry Mass Formula Based on omega- phi Mixing,''
Phys. Rev. \textbf{132}, 434-441 (1963)
doi:10.1103/PhysRev.132.434

\bibitem{Coleman1964}
S.~Coleman and H.~J.~Schnitzer,
``Departures from the Eightfold Way. 2. Baryon Electromagnetic Masses,''
Phys. Rev. \textbf{136}, B223-B226 (1964)
doi:10.1103/PhysRev.136.B223

\bibitem{Schwinger:1957em}
J.~S.~Schwinger,
``A Theory of the Fundamental Interactions,''
Annals Phys. \textbf{2}, 407-434 (1957)
doi:10.1016/0003-4916(57)90015-5

\bibitem{christlee75}
N.~H.~Christ and T.~D.~Lee,
``Quantum Expansion of Soliton Solutions,''
Phys. Rev. D \textbf{12} (1975), 1606
doi:10.1103/PhysRevD.12.1606

\bibitem{goldstone}
J.~Goldstone and R.~Jackiw,
``Quantization of Nonlinear Waves,''
Phys. Rev. D \textbf{11} (1975), 1486-1498
doi:10.1103/PhysRevD.11.1486


%\bibitem{mak78}
%V.~G.~Makhankov,
%``Dynamics of Classical Solitons In Nonintegrable Systems,''
%Phys. Rept. \textbf{35} (1978), 1-128
%doi:10.1016/0370-1573(78)90074-1

%\bibitem{moshir81}
%M.~Moshir,
%``Soliton - Anti-soliton Scattering and Capture in $\lambda \phi^4$ Theory,''
%Nucl. Phys. B \textbf{185} (1981), 318-332
%doi:10.1016/0550-3213(81)90320-5

%\bibitem{adamwall}
%C.~Adam, K.~Oles, T.~Romanczukiewicz and A.~Wereszczynski,
%``Spectral Walls in Soliton Collisions,''
%Phys. Rev. Lett. \textbf{122} (2019) no.24, 24160170.
%doi:10.1103/PhysRevLett.122.241601
%[arXiv:1903.12100 [hep-th]].

%\bibitem{lankink}
%Y.~Zhong, X.~L.~Du, Z.~C.~Jiang, Y.~X.~Liu and Y.~Q.~Wang,
%``Collision of two kinks with inner structure,''
%JHEP \textbf{02}, 153 (2020)
%doi:10.1007/JHEP02(2020)153
%[arXiv:1906.02920 [hep-th]].

%\bibitem{campres}
%D.~K.~Campbell, J.~F.~Schonfeld and C.~A.~Wingate,
%``Resonance Structure in Kink - Antikink Interactions in $\phi^{4}$ Theory,''
%Physica D \textbf{9} (1983), 1
%FERMILAB-PUB-82-051-THY.

%\bibitem{alonso2020}
%A.~Alonso-Izquierdo, L.~M.~Nieto and J.~Queiroga-Nunes,
%``Scattering between wobbling kinks,''
%[arXiv:2007.15517 [hep-th]].

%\bibitem{takyi}
%I.~Takyi and H.~Weigel,
%``Collective Coordinates in One-Dimensional Soliton Models Revisited,''
%Phys. Rev. D \textbf{94} (2016) no.8, 085008
%doi:10.1103/PhysRevD.94.085008
%[arXiv:1609.06833 [nlin.PS]].

%\bibitem{quintero2000}
%N.~R.~Quintero, A.~Sanchez and F.~G.~Mertens,
%``Resonances in the dynamics of $\phi^4$ kinks perturbed by ac forces,''
%Phys. Rev. E \textbf{62} (2000), 5695-5705
%doi:10.1103/PhysRevE.62.5695
%[arXiv:cond-mat/0006313 [cond-mat.stat-mech]].

\bibitem{dhn2}
  R.~F.~Dashen, B.~Hasslacher and A.~Neveu,
  ``Nonperturbative Methods and Extended Hadron Models in Field Theory 2. Two-Dimensional Models and Extended Hadrons,''
  Phys.\ Rev.\ D {\bf 10} (1974) 4130.
 doi:10.1103/PhysRevD.10.4130


\bibitem{me2looplett}
J.~Evslin and H.~Guo,
``An Alternative to Collective Coordinates,''
Accepted for publication in PRD,
[arXiv:2101.08028 [hep-th]].



\bibitem{memassa}
J.~Evslin,
``Well-defined quantum soliton masses without supersymmetry,''
Phys. Rev. D \textbf{101} (2020) no.6, 065005
doi:10.1103/PhysRevD.101.065005
[arXiv:2002.12523 [hep-th]].


\bibitem{me2loop}
J.~Evslin and H.~Guo,
``Two-Loop Scalar Kinks,''
[arXiv:2012.04912 [hep-th]].

\bibitem{rajaramanb}
  Rajaraman ~ R.
  ``Solitons and Instantons,''
 Physics Bulletin 34.1(1983):29-29.

\bibitem{rajaraman}
  R.~Rajaraman,
  ``Some Nonperturbative Semiclassical Methods in Quantum Field Theory: A Pedagogical Review,''
  Phys.\ Rept.\  {\bf 21} (1975) 227.
  doi:10.1016/0370-1573(75)90016-2

\bibitem{mekink}
J.~Evslin,
``Manifestly Finite Derivation of the Quantum Kink Mass,''
JHEP \textbf{11} (2019), 161
doi:10.1007/JHEP11(2019)161
[arXiv:1908.06710 [hep-th]].

\bibitem{me1loop}
H.~Guo and J.~Evslin,
``Finite derivation of the one-loop sine-Gordon soliton mass,''
JHEP \textbf{02}, 140 (2020)
doi:10.1007/JHEP02(2020)140
[arXiv:1912.08507 [hep-th]].

%\bibitem{baiyang}
%J. Evslin and B. Zhang, In Preparation.

\bibitem{rebhan}
  A.~Rebhan and P.~van Nieuwenhuizen,
  ``No saturation of the quantum Bogomolnyi bound by two-dimensional supersymmetric solitons,''
  Nucl.\ Phys.\ B {\bf 508} (1997) 449
  doi:10.1016/S0550-3213(97)00625-1, 10.1016/S0550-3213(97)80021-1
 [hep-th/9707163].

%\bibitem{lit}
%A.~Litvintsev and P.~van Nieuwenhuizen,
%``Once more on the BPS bound for the SUSY kink,''
%[arXiv:hep-th/0010051 [hep-th]].

\bibitem{local}
A.~S.~Goldhaber, A.~Litvintsev and P.~van Nieuwenhuizen,
``Local Casimir energy for solitons,''
Phys. Rev. D \textbf{67} (2003), 105021
doi:10.1103/PhysRevD.67.105021
[arXiv:hep-th/0109110 [hep-th]].

%\bibitem{nastase}
%H.~Nastase, M.~A.~Stephanov, P.~van Nieuwenhuizen and A.~Rebhan,
%``Topological boundary conditions, the BPS bound, and elimination of ambiguities in the quantum mass of solitons,''
%Nucl. Phys. B \textbf{542} (1999), 471-514
%doi:10.1016/S0550-3213(98)00773-1
%[arXiv:hep-th/9802074 [hep-th]].


\bibitem{cahill76}
K.~E.~Cahill, A.~Comtet and R.~J.~Glauber,
``Mass Formulas for Static Solitons,''
Phys. Lett. B \textbf{64} (1976), 283-285
doi:10.1016/0370-2693(76)90202-1


\bibitem{mewick}
J.~Evslin,
``Normal Ordering Normal Modes,''
Eur. Phys. J. C \textbf{81} (2021) no.1, 92
doi:10.1140/epjc/s10052-021-08890-7
[arXiv:2007.05741 [hep-th]].

\bibitem{me2stato}
J.~Evslin,
``Constructing Quantum Soliton States Despite Zero Modes,''
[arXiv:2006.02354 [hep-th]].

%\bibitem{lsz77}
%A.~Rouet and K.~Yoshida,
%``Scattering Theory in the Presence of a Soliton Asymptotic Theory,''
%Phys. Lett. B \textbf{70} (1977), 117-119
%doi:10.1016/0370-2693(77)90358-6

%\bibitem{lsz88}
%K.~A.~Sveshnikov,
%``FERMIONIC AND BOSONIC SCATTERING PHASES ON A TOPOLOGICAL KINK,''
%Phys. Lett. A \textbf{134} (1988), 47-53
%doi:10.1016/0375-9601(88)90545-2

%\bibitem{dorey93}
%N.~Dorey, M.~P.~Mattis and J.~Hughes,
%``Soliton quantization and internal symmetry,''
%Phys. Rev. D \textbf{49} (1994), 3598-3611
%doi:10.1103/PhysRevD.49.3598
%[arXiv:hep-th/9309018 [hep-th]].

%\bibitem{dorey94}
%N.~Dorey and T.~J.~Hollowood,
%``Quantum scattering of charged solitons in the complex sine-Gordon model,''
%Nucl. Phys. B \textbf{440} (1995), 215-236
%doi:10.1016/0550-3213(95)00074-3
%[arXiv:hep-th/9410140 [hep-th]].

%\bibitem{babulsz}
%H.~M.~Babujian, A.~Fring, M.~Karowski and A.~Zapletal,
%``Exact form-factors in integrable quantum field theories: The Sine-Gordon model,''
%Nucl. Phys. B \textbf{538} (1999), 535-586
%doi:10.1016/S0550-3213(98)00737-8
%[arXiv:hep-th/9805185 [hep-th]].

%\bibitem{adammultwall}
%C.~Adam, K.~Oles, T.~Romanczukiewicz and A.~Wereszczynski,
%``Kink-antikink scattering in the $\phi^4$ model without static intersoliton forces,''
%Phys. Rev. D \textbf{101} (2020) no.10, 105021
%doi:10.1103/PhysRevD.101.105021
%[arXiv:1909.06901 [hep-th]].

\bibitem{gjs}
J.~L.~Gervais, A.~Jevicki and B.~Sakita,
``Perturbation Expansion Around Extended Particle States in Quantum Field Theory. 1.,''
Phys. Rev. D \textbf{12} (1975), 1038
doi:10.1103/PhysRevD.12.1038

\bibitem{jackiw77}
R.~Jackiw,
``Quantum Meaning of Classical Field Theory,''
Rev. Mod. Phys. \textbf{49} (1977), 681-706
doi:10.1103/RevModPhys.49.681



\end{thebibliography}
\end{document}